\documentclass[aps,prl,twocolumn,groupedaddress,amsmath,amssymb]{revtex4}

\newcommand{\pat}{\partial}

\def\beq{\begin{equation}}
\def\eeq{\end{equation}}

\newcommand{\christoffel}[3]{\genfrac{\{}{\}}{0pt}{}{#1\hfill}{#2 #3}}

\begin{document}

\title{Relationship of gauge gravitation theory in Riemann-Cartan spacetime and general relativity theory
}

\author{A. V. Minkevich}

\email[]{MinkAV@bsu.by; awm@matman.uwm.edu.pl}

%\address{Belarusian State University, F.Skariny Av., 4, 220050 Minsk,
%Republic of Belarus}

 \affiliation{Department of Theoretical Physics and Astrophysics, Belarussian State University,
Minsk, Belarus}
 \affiliation{ Department of Physics and Computer Methods, Warmia
and Mazury University in Olsztyn, Poland}

%\date{\today}

\begin{abstract}
The simplest variant of gauge gravitation theory in Riemann-Cartan spacetime
leading to the solution of the problem of cosmological singularity and dark
energy problem is investigated. It is shown that this theory by certain
restrictions on indefinite parameters of gravitational Lagrangian in the case
of usual gravitating systems leads to Einstein gravitational equations with
effective cosmological constant.
\end{abstract}

\maketitle

\section{Introduction}

The investigation of gauge gravitation theory in Riemann-Cartan spacetime
(GTRC), which is a necessary generalization of metric gravitation theory in the
framework of gauge approach by including the Lorentz group into the gauge group
corresponding to gravitational interaction, shows that GTRC allows to solve
some principal problems of general relativity theory (GR) by virtue of the
change of gravitational interaction by certain physical conditions in the frame
of GTRC in comparison with GR (see for example \cite{b1,b2,b3}). The change of
gravitational interaction is provoked by more complicated structure of physical
spacetime, namely by spacetime torsion. In the frame of GTRC the gravitational
interaction in the case of usual gravitating matter with positive values of
energy density and pressure can be repulsive. The effect of gravitational
repulsion appears at extreme conditions when energy density and pressure are
extremely high and also in case when energy density is very small and vacuum
effect of gravitational repulsion is essential. This allows to solve the
problem of cosmological singularity and to explain accelerating cosmological
expansion at present epoch without using the notion of dark energy. Given data
were obtained by study of isotropic cosmology built in the frame of GTRC based
on general expression of gravitational Lagrangian including both a scalar
curvature and quadratic in the curvature and torsion invariants with indefinite
parameters by certain restrictions on these parameters. A physical cause of a
change of gravitational interaction in GTRC is connected with the fact that
torsion according to gravitational equations is function of energy density and
pressure and together with energy-momentum tensor affects on spacetime metric.
The torsion plays the principal role at extreme conditions by formation of
limiting energy density for gravitating matter \cite{b4} and also leads to
formation of effective cosmological constant at asymptotics of cosmological
models by virtue of influence on physical spacetime in the vacuum having the
structure of Riemann-Cartan continuum with de Sitter metric (but not Minkowski
spacetime) \cite{b5}.

The following question appears: what is possible role of the torsion in
astrophysics in the case of usual gravitating systems, for which energy density
is much smaller than limiting energy density \footnote{We don't discuss here
massive stars collapsing in GR with formation of singular black holes. In the
frame of GTRC such objects are impossible if limiting energy density exists in
the nature.} but greater than average energy density in the Universe at present
epoch (or effective cosmological constant). It should be noted that one torsion
function at asymptotics of cosmological models has the structure (see below)
which can be essential quantitatively in newtonian approximation though the
evolution of cosmological models at asymptotics coincides practically with that
of Friedmann cosmological models with cosmological constant. The study of this
question is particular case of research of relationship of GTRC and GR.

This paper is devoted to investigation of relationship of GR and the simplest
GTRC (minimum GTRC) which allows to build the theory of regular accelerating
Universe.

\section{Isotropic cosmology and minimum gauge gravitation theory in Riemann-Cartan spacetime}

In the beginning lets introduce the base definitions and relations used in this
paper. In the framework of GTRC the role of gravitational field variables play
the orthonormalized tetrad $h^i{}_\mu$ and the Lorentz connection
$A^{ik}{}_\mu$; corresponding field strengths are the torsion tensor
$S^i{}_{\mu\nu}$ and the curvature tensor $F^{ik}{}_{\mu\nu}$ defined as
\[
S^i{}_{\mu \,\nu }  = \partial _{[\nu } \,h^i{}_{\mu ]}  - h_{k[\mu }
A^{ik}{}_{\nu ]}\,,
\]
\[
F^{ik}{}_{\mu\nu }  = 2\partial _{[\mu } A^{ik}{}_{\nu ]}  + 2A^{il}{}_{[\mu }
A^k{}_{|l\,|\nu ]},\
\]
where holonomic and anholonomic spacetime coordinates are denoted by means of
greek and latin indices respectively \footnote{Like our previous papers we will
use notations corresponding to the following relation between holonomic
connection $\Gamma^{\lambda}{}_{\mu\nu }$ and $A^{ik}{}_\mu$:
$\Gamma^{\lambda}{}_{\mu\nu } = h_i{}^{\lambda } (\partial _{\nu } \,h^i{}_{\mu
} - h_{k\mu } A^{ik}{}_{\nu })$. Then the tensor $F^{\rho}{}_{\sigma\mu\nu
}=h_i{}^{\rho} h_{k \sigma } F^{ik}{}_{\mu\nu }= 2\partial _{[\nu
}\Gamma^{\rho}{}_{|\sigma\,|\mu] }+ 2 \Gamma^{\rho}{}_{\lambda [\nu }
\Gamma^{\lambda}{}_{|\sigma\,| \mu] }$ has the opposite sign as compared with
frequently defining curvature tensor and $S^{\lambda}{}_{\mu \,\nu }
=\Gamma^{\lambda}{}_{[\mu\nu] }$ (cf.\cite{b7,b8}). The signature of spacetime
metric is (-2).}. Isotropic cosmology in Riemann-Cartan spacetime
investigated in a number of papers (see for example \cite{b1,b2,b3,b4,b5,b6})
was built by using the gravitational Lagrangian given in the following
sufficiently general form
\begin{eqnarray}\label{1}%\fl
\mathcal{L}_{\rm g}=  f_0\,
F+F^{\alpha\beta\mu\nu}\left(f_1\:F_{\alpha\beta\mu\nu}+f_2\:
F_{\alpha\mu\beta\nu}+f_3\:F_{\mu\nu\alpha\beta}\right)  \nonumber \\
+ F^{\mu\nu}\left(f_4\:F_{\mu\nu}  +f_5\: F_{\nu\mu}\right)
+ f_6\:F^2 \nonumber \\
+S^{\alpha\mu\nu}\left(a_1\:S_{\alpha\mu\nu}+a_2\: S_{\nu\mu\alpha}\right)
+a_3\:S^\alpha{}_{\mu\alpha}S_\beta{}^{\mu\beta},
\end{eqnarray}
where $F_{\mu\nu}=F^{\alpha}{}_{\mu\alpha\nu}$, $F=F^\mu{}_\mu$, $f_i$
($i=1,2,\ldots,6$), $a_k$ ($k=1,2,3$) are indefinite parameters, $f_0=(16\pi
G)^{-1}$, $G$ is Newton's gravitational constant (the light speed in the vacuum
$c=1$). Gravitational equations of PGTG obtained from the action integral $ I =
\int {\left( {{\cal L}_g + {\cal L}_m } \right)\,}h d^4 x$, where
$h=\det{\left(h^i{}_\mu\right)}$ and ${\cal L}_m$ is the Lagrangian of
gravitating matter, contain the system of 16+24 equations corresponding to
gravitational variables $h^i{}_\mu$ and $A^{ik}{}_\mu$. By using minimal
coupling of gravitational field with matter, the energy-momentum tensor
$T_{i}{}^{\mu}=-\frac {1}{h} \frac{\delta{\cal L}_m} {\delta h^{i}{}_{\mu}}$
and spin momentum tensor $J_{[ik]}{}^{\mu}=-\frac {1}{h} \frac{\delta{\cal
L}_m} {\delta A^{ik}{}_{\mu}}$ of gravitating matter manifest as sources of
gravitational field in gravitational equations. Gravitational equations are
complicated system of differential equations in partial derivatives with
indefinite parameters $f_i$ and $a_k$. Physical consequences depend essentially
on restrictions on these parameters. Some of such restrictions were obtained by
investigation of isotropic cosmology, notably the solution of cosmological
problems mentioned previously was obtained by the following restrictions: $2a_1
+ a_2  + 3a_3=0$ and $2f_1 - f_2=0$. Then cosmological equations and equations
for torsion functions include three indefinite parameters: parameter
$\alpha=\frac{f}{3f_0^2}$  ($f = f_1  + \frac{{f_2 }} {2} + f_3 + f_4 + f_5 +
3f_{6}>0$) with inverse dimension of energy density, parameter $b=a_2 - a_1$
with the same dimension as $f_0$ and dimensionless parameter $\omega= \frac
{f_2 + 4f_3 + f_4 + f_5} {f}$. The correspondence of GTRC to GR in linear
approximation with respect to metric and torsion takes place, if we suppose
that the value of $\alpha^{-1}$ corresponds to the scale of high
energy densities \cite{b1} and the parameter $\omega$ is small $0 < \omega\ll
1$. In addition the correspondence of GTRC to GR leads to the following
condition for parameter $b$: $0<x=1-\frac{b}{f_0}\ll 1$ (see below). Besides
the scale of high energy densities defined by parameter $\alpha^{-1}$
there is the second scale of extremely high energy densities of order $(\omega
\alpha)^{-1}$, which determines the value of limiting energy density.

Any homogeneous isotropic model (HIM) is described by three functions of time:
the scale factor of Robertson-Walker metric $R(t)$ and two torsion functions
$S_{1}(t)$ and $S_{2}(t)$. Cosmological equations generalizing Friedmann
cosmological equations of GR take the form
\begin{eqnarray}\label{2.2}%\fl
    \frac{k}{R^2} + (H-2S_1)^2 -S_2^2= \nonumber\\
    \frac{1}{{6f_0 Z}}
        \left[
            {\rho  -6 b S_2^2
            + \frac{\alpha }{4} \left( {\rho  - 3p - 12bS_2^2 } \right)^2 }
        \right],
\end{eqnarray}
\begin{eqnarray}\label{2.3}%\fl
    \dot{H}-2\dot{S}_1 +H (H-2S_1)= \nonumber\\
    -\frac{1} {{12f_0 Z}}
        \left[
            \rho  + 3p - \frac{\alpha } {2} \left( {\rho  - 3p - 12bS_2^2 } \right)^2
        \right],
\end{eqnarray}
where $H=\dot{R}/R $ is the Hubble parameter (a dot denotes the differentiation
with respect to time), $k=+1,0,-1$ for closed, flat and open models
respectively, $\rho$ is energy density, $p$ is pressure and $Z=1+\alpha\left(
\rho - 3p - 12b S_2^2\right)$. The torsion functions $S_1$ and $S_2$ are
\begin{eqnarray}\label{2.4}%\fl
    S_1  = -\frac{\alpha }{4Z} [\dot \rho
    - 3 \dot p + 12f_0 \omega H S_2^2
    -12( {2b - \omega f_0 } ) S_2 \dot S_2].
\end{eqnarray}
\begin{eqnarray}\label{2.5}
 S_{2}^{2}  = \frac{\rho - 3p}{12b} + \frac
{1-(b/2f_0) (1 +  \sqrt{X})} {12b \alpha (1- \omega/4)},
\end{eqnarray}
where
\begin{equation}\label{2.6}
X=1+ \omega (f_0^2/b^2) [1- (b/f_0) - 2(1- \omega /4) \alpha ( \rho + 3p)]\ge
0.
\end{equation}
As consistent with cosmological equations the energy density $\rho$ and
pressure $p$ satisfy the equation:
\begin{equation}\label{2.7}
\dot{\rho}+3H\left(\rho+ p\right)=0.
\end{equation}

The torsion function $S_2$ plays important role at asymptotics when energy
density is sufficiently small: $\alpha (\rho+3p)\ll1$. Then according to
(5)-(6) if $0<x=1-\frac{b}{f_0}\ll 1$ we have in the lowest approximation
with respect to $x$:
\begin{equation}\label{2.8}
S_2^2  = \frac {1} {12b} \left[\rho  - 3p + \frac {1  - b/f_0} {\alpha}\right],
\end{equation}
The presence of constant term in (8) leads to appearance of effective
cosmological constant in cosmological equations, which at asymptotics take the
form:
\begin{equation}\label{2.9}
    \frac{k}{R^2 } + H^2  = \frac{1}{6f_0 }\left[\rho \frac{f_0}{b} + \frac{1}{4\alpha} \left(1 - \frac{b}{f_0}\right)^2
    \frac{f_0}{b} \right],
\end{equation}
\begin{equation}\label{2.10}
    \dot H + H^2  =  - \frac{1} {{12f_0 }}\left[ (\rho + 3p) \frac{f_0}{b} - \frac{1}{2\alpha}
    \left(1 - \frac{b}{f_0}\right)^2 \frac{f_0}{b}\right].
\end{equation}
In situation when the value of energy density $\rho$ is comparable with
effective cosmological constant,  equations (9)-(10) practically coincide
with Friedmann cosmological equations with cosmological constant. However, in
situation when effective cosmological constant in cosmological equations
(9)-(10) can be neglected in comparison with energy density $\rho$, the
evolution of HIM described by (9)-(10) weakly differs from that of
Friedmann cosmological equations because we have the term $\rho \frac{f_0}{b}
\approx \rho (1+x)$ instead $\rho$ in right part of (9)-(10).

The dependence of $S_2^2$ on energy density and pressure is similar to that of
energy-momentum tensor in gravitational equations, moreover the constant term
in expression (8) is much greater than effective cosmological constant in
cosmological equations. However, terms in (2)-(3) depending on $S_2^2$ are
mutually cancelled and influence of torsion function $S_2$ appears by formation
of effective cosmological constant because of terms $S_2^4$ in cosmological
equations \footnote{This means that establishing of correspondence between GTRC
and GR in linear approximation with respect to torsion \cite{b7,b1} is not
sufficient.}. Is it distinctive feature of HIM connected probably with its high
symmetry or is it some characteristic property of gravitational equations of
GTRC?

We will study this question by using discussed earlier restrictions on
indefinite parameters introduced in the frame of isotropic cosmology. The
remaining indefinite parameters in gravitational Lagrangian (1) can be
excluded by using additional physical considerations. So we can use
restrictions on indefinite parameters obtained in \cite{b7} from analysis of
particle content of GTRC in linear approximation and exception of ghosts and
tachyons. Restrictions on indefinite parameters obtained in the frame of
isotropic cosmology are compatible with the following conditions:
$f_1=f_2=f_3=f_4=0$ and
\begin{eqnarray}\label{2.11}
a_1=f_0 (1-x),\qquad a_2=2f_0 (1-x),
\nonumber \\
 a_3=-\frac{4}{3}f_0(1-x), \qquad f_5=3f_0^2 \alpha \omega,
\nonumber \\
  f_6=f_0^2 \alpha (1-\omega) \qquad (x=1- \frac{b} {f_0}).
\end{eqnarray}
The particle content of GTRC with such restrictions on indefinite parameters
includes besides massless graviton massive particles with spin-parity $2^{+}$.
\footnote{The strict analysis of particle content has to be connected
with consideration of gravitational perturbations above the vacuum spacetime
having the structure of Riemann-Cartan continuum with de Sitter metric.
However, the deviation of the structure of the vacuum space-time from Minkowski
space-time, which is essential at cosmological scale, can be unimportant by
local analysis given in \cite{b7} because of smallness of values of parameter
$H$ and torsion for the vacuum. In addition the conditions of particle content
of minimum GTRC noted above are valid approximately by virtue of restriction $0
< \omega\ll 1$.}. Our further consideration will be connected with this GTRC - so-called minimum
GTRC.

\section{Relationship of minimum GTRC and GR}

Gravitational equations of minimum GTRC have the following form:
\begin{eqnarray}\label{3.1}
\nabla_{\nu}U_{i}{}^{\mu\nu}+2S^k{}_{i\nu}U_k{}^{\mu\nu}
\nonumber\\
+2(f_0+2f_6\:F)F^{\mu}{}_i
+2f_5(F_{ki}F^{\mu
k}+F^{\mu}{}_{kim}F^{mk})
\nonumber\\
- h_{i}{}^{\mu}(f_0 F+ f_5 F^{\mu\nu} F_{\nu\mu} +
f_6\:F^2 +
\nonumber\\
S^{\alpha\mu\nu}\left(a_1\:S_{\alpha\mu\nu}+a_2\:
S_{\nu\mu\alpha}\right) +a_3\:S^\alpha{}_{\mu\alpha}S_\beta{}^{\mu\beta})
\nonumber\\
=-T_{i}{}^{\mu},
\end{eqnarray}
\begin{eqnarray}\label{3.2}
4\nabla_{\nu}[(f_0/2+f_6\:F)h_{[i}{}^{\nu}h_{k]}{}^{\mu}+
+f_5\:F^{[\mu}{}_{[k}h_{i]}{}^{\nu]}]+U_{[ik]}{}^{\mu}
\nonumber\\
=-J_{[ik]}{}^{\mu},
\end{eqnarray}
where
$U_{i}{}^{\mu\nu}=2(a_1\:S_{i}{}^{\mu\nu}-a_2\:S^{[\mu\nu]}{}_{i}-a_3\:S_{\alpha}{}^{\alpha
[\mu }h_{i}{}^{\nu]})$, $\nabla_{\nu}$ denotes the covariant operator having
the structure of the covariant derivative defined in the case of tensor
holonomic indices by means of Christoffel coefficients
$\christoffel{\lambda}{\mu}{\nu}$ and in the case of tetrad tensor indices by
means of anholonomic Lorentz connection $A^{ik}{}_{\nu}$ (for example
$\nabla_{\nu} h^{i}{}_{\mu}=\partial _{\nu } \,h^i{}_{\mu
}-\christoffel{\lambda}{\mu}{\nu}\, h^{i}{}_{\lambda}-A^{ik}{}_{\nu}h_{k\mu}$).
Analytic analysis is possible if $\omega\ll 1$, then according to (11) $f_5
\ll f_6$. We will analyze the system of equations (12)-(13) in the case of
spinless matter ($J_{[ik]}{}^{\mu}=0$) by neglecting terms with $f_5$. Then the
system of gravitational equations takes the form:
\begin{eqnarray}\label{3.3}
\nabla_{\nu}U_{i}{}^{\mu\nu}+2S^k{}_{i\nu}U_k{}^{\mu\nu}+
2(f_0+2f_6\:F)F^{\mu}{}_i -
\nonumber\\
h_{i}{}^{\mu}(f_0 F+ f_6\:F^2
+S^{\alpha\mu\nu}\left(a_1\:S_{\alpha\mu\nu}+a_2\: S_{\nu\mu\alpha}\right)
\nonumber\\
+a_3\:S^\alpha{}_{\mu\alpha}S_\beta{}^{\mu\beta}) =-T_{i}{}^{\mu},
\end{eqnarray}
\begin{eqnarray}\label{3.4}
4\nabla_{\nu}[(f_0/2+f_6\:F)h_{[i}{}^{\nu}h_{k]}{}^{\mu} ]+U_{[ik]}{}^{\mu}=0.
\end{eqnarray}
By taking into account that
\begin{equation}\label{3.5}
\Gamma^{\lambda}{}_{\mu\nu }= \christoffel{\lambda}{\mu}{\nu} +
K^{\lambda}{}_{\mu\nu },
\end{equation}
where
\begin{equation}\label{3.6}
K^{\lambda}{}_{\mu\nu }= S^{\lambda}{}_{\mu\nu } + S_{\mu\nu }{}^{\lambda} +
S_{\nu\mu }{}^{\lambda},
\end{equation}
we obtain that
\begin{equation}\label{3.7}
\nabla_{\nu} h_{i}{}^{\mu}=\partial _{\nu } \,h_i{}^{\mu
}+\christoffel{\mu}{\lambda}{\nu}\, h_{i}{}^{\lambda}+A^{k}{}_{i
\nu}h_{k}{}^{\mu}= - K^{\mu}{}_{\lambda\nu } h_i{}^{\lambda}.
\end{equation}
By using (18) and the relation $a_2=2a_1$ and by multiplying eq. (15) with
$h^{i}{}_{\sigma} h^{k}{}_{\rho}$ we transform the equation (15) to the
following form:
\begin{eqnarray}\label{3.4}
2f_0 (1 + \frac{2f_6} {f_0} F) (S^{\mu}{}_{\sigma\rho } + 2S^{\nu}{}_{\nu
[\sigma}{}{\delta}^{\mu}_{\rho]}) - a_2 S^{\mu}{}_{\sigma\rho }
 \nonumber\\
 + a_3 S^{\nu}{}_{\nu [\sigma}{}{\delta}^{\mu}_{\rho]} + 4f_6 {\pat}_{\nu}F
{\delta}^{\nu}_{[\sigma}{\delta}^{\mu}_{\rho]}=0.
\end{eqnarray}
By denoting $Z_1=1 + \frac{2f_6} {f_0} F \approx 1 + 2f_0 {\alpha} F$ we write
eq. (19) in the form\footnote{The quantity $Z_1$ corresponds to $Z$ used
earlier in cosmology.}:
\begin{equation}\label{3.9}
 (2f_0 Z_1 - a_2) S^{\mu}{}_{\sigma\rho } +  (4f_0 Z_1 + a_3)
 S^{\nu}{}_{\nu [\sigma}{}{\delta}^{\mu}_{\rho]}+ 4f_6 {\pat}_{\nu}F
{\delta}^{\nu}_{[\sigma}{\delta}^{\mu}_{\rho]}=0.
\end{equation}
From (20) follows that if there is nonvanishing component of torsion with
${\mu} \neq {\rho}$ and ${\mu} \neq {\sigma}$ we obtain $f_0 Z_1 -\frac {a_2}
{2}=0$ and for minimum GTRC with restrictions (11) $Z_1=\frac{b}{f_0}$. As
result the scalar curvature $F$ is constant: $F=-\frac{1-\frac{b}{f_0}}{2f_0
{\alpha}}$. In the case ${\mu}={\rho}$ and ${\mu} \neq {\sigma}$ eq. (20)
leads to $S^{\nu}{}_{\nu \sigma}=0$.

Now we will analyze gravitational equation (14). By using that $\nabla_{\nu}
h^{i}{}_{\mu}=K^{\lambda}{}_{\mu\nu } h^i{}_{\lambda}$ and by multiplying eq.
(14) with $h^{i}{}_{\lambda}$ we transform eq. (14) to the following form:
\begin{eqnarray}\label{3.10}
\nabla_{\nu}U_{\lambda}{}^{\mu\nu}- K^{\rho}{}_{\lambda\nu }
U_{\rho}{}^{\mu\nu} +2S^{\rho}{}_{\lambda\nu}U_{\rho}{}^{\mu\nu}
\nonumber\\
+ 2f_0(1+\frac {2f_6}{f_0}\:F)F^{\mu}{}_{\lambda} -
\delta^{\mu}_{\lambda}(f_0 F+ f_6\:F^2
\nonumber\\
+S^{\alpha\mu\nu}\left(a_1\:S_{\alpha\mu\nu}+a_2\: S_{\nu\mu\alpha}\right)
+a_3\:S^\alpha{}_{\mu\alpha}S_\beta{}^{\mu\beta}) =-T_{\lambda}{}^{\mu}.
\end{eqnarray}
From (21) we obtain the following expression for scalar curvature:
\begin{equation}\label{3.11}
 F= \frac {1}{2f_0} [T- 2b S_{\lambda\mu\nu }(S^{\lambda\mu\nu }- 2
 S^{\mu\nu\lambda})],
\end{equation}
where $T=T_{\mu}{}^{\mu}$. The expression (22) corresponds to scalar
curvature of HIM obtained earlier \cite{b1}.

By using obtained value of constant scalar curvature
$F=-\frac{1-\frac{b}{f_0}}{2f_0 {\alpha}}$ and the formula $S^{\nu}{}_{\nu
\sigma}=0$ we transform eq. (21) to the form:
\begin{eqnarray}\label{3.12}
\frac{1}{2b} (\nabla_{\nu}U_{\lambda}{}^{\mu\nu}- K^{\rho}{}_{\lambda\nu }
U_{\rho}{}^{\mu\nu} +2S^{\rho}{}_{\lambda\nu}U_{\rho}{}^{\mu\nu})+
F^{\mu}{}_{\lambda} -\frac{1}{2} \delta^{\mu}_{\lambda} F-
\nonumber\\
\frac{1}{2}\delta^{\mu}_{\lambda} S^{\alpha\mu\nu}(S_{\alpha\mu\nu}+
2S_{\nu\mu\alpha}) =-\frac{1}{2b} (T_{\lambda}{}^{\mu} + \delta^{\mu}_{\lambda}
\frac{(1-\frac{b}{f_0})^2}{12\alpha}).
\end{eqnarray}
The tensor $F^{\rho}{}_{\sigma\mu\nu }$ can be presented in the form of sum of
riemannian part depending on Christoffel coefficients and denoting by
$R^{\rho}{}_{\sigma\mu\nu }(\christoffel{}{}{})$ and part depending on torsion
and denoting by $F^{\rho}{}_{\sigma\mu\nu }(K)$. As result the tensors
$F^{\mu}{}_{\lambda}$ and $F$ in (23) are divided by the following way:
$F^{\mu}{}_{\lambda}= R^{\mu}{}_{\lambda}+ F^{\mu}{}_{\lambda}(K)$ and
$F=R+F(K)$ and by taking into account $S^{\nu}{}_{\nu \sigma}=0$ we have:
\begin{eqnarray}\label{3.12}
F^{\mu}{}_{\lambda}(K)=-\nabla_{\nu}K^{\nu\mu}{}_{\lambda}+K_{\nu\rho\lambda}
K^{\rho\mu\nu} ,
\nonumber\\
\qquad  F(K)=K_{\lambda\mu\nu} K^{\mu\nu\lambda}.
\end{eqnarray}
By using formulas (24) and (17) we find that all terms with torsion in
(23) are mutually eliminated and eq. (23) takes the form of Einstein
gravitational equations with cosmological constant:
\begin{equation}\label{3.14}
R^{\mu}{}_{\lambda} -\frac{1}{2} \delta^{\mu}_{\lambda} R=-\frac{1}{2b}
(T_{\lambda}{}^{\mu} + \delta^{\mu}_{\lambda}
\frac{(1-\frac{b}{f_0})^2}{12\alpha}).
\end{equation}
Besides effective cosmological constant the influence of torsion in eq. (25)
appears via the change of gravitational constant, however, because the value of
$b$ is very near to $f_0$ corresponding consequences are insignificant. Note
that analysis leading to eq. (25) is not applicable at extreme conditions
near limiting energy density, where terms with parameter $\omega$ in
gravitational equations play principal role.

It should be noted that equations of minimum GTRC (12)-(13) like GTRC based
on gravitational Lagrangian (1) have a number of solutions which are
unacceptable from physical point of view. In particular, as it was shown in
\cite{b7} any vacuum solution of GR with vanishing torsion is exact solution of
GTRC independently on values of indefinite parameters $f_i$ and $a_k$ while
solutions of GTRC far from spatially limited systems have to tend to the vacuum
solution with nonvanishing torsion. In connection with this we have to state
the criterion \cite{b1}, which allows to distinguish acceptable solutions from
unphysical ones. Such criterion can be based on investigation of solutions at
asymptotics: far from spatially limited systems and at asymptotics of
cosmological models solutions of GTRC have to tend to the vacuum solution in
the form of corresponding Riemann-Cartan continuum.

\section{Conclusion}

We obtain that minimum GTRC by neglecting spin effects does not lead to
essential distinction in behavior of usual astrophysical objects in comparison
with GR, if our assumption ($\omega \ll 1$) is valid. Gravitational equations
(25) were derived in the case of spinless matter. Investigation of spin effects
in the frame of minimum GTRC similar to that in Einstein-Cartan theory
\cite{b9} is of physical interest. The interaction of proper angular moments of
astrophysical objects with spacetime torsion can have the important physical
significance.

\label{lastpage-01}
\end{document}